\documentclass[aps,prl,reprint,longbibliography,superscriptaddress]{revtex4-2}

\usepackage{amsmath,amssymb}
\usepackage{graphicx}
\usepackage{xcolor}
\usepackage{mathrsfs}
\usepackage{stmaryrd}
\usepackage{hyperref}
\usepackage{enumitem}
\usepackage{accents}
\hypersetup{colorlinks,linkcolor={blue!80!black},citecolor={blue!80!black},urlcolor={blue!80!black}}

\renewcommand{\pi}{\uppi}

\newcommand{\figref}[2]{[Fig.~\hyperref[#1]{\ref*{#1}(#2)}]}
\newcommand{\figrefi}[2]{[Fig.~\hyperref[#1]{\ref*{#1}(#2)}, inset]}
\newcommand{\textfigref}[2]{Fig.~\hyperref[#1]{\ref*{#1}(#2)}}

\newcommand{\figrefp}[2]{\hyperref[#1]{\ref*{#1}(#2)}}

\begin{document}

\title{Comment on ``Faceting and Flattening of Emulsion Droplets: A Mechanical Model''}
\author{Pierre A. Haas}
\email{haas@pks.mpg.de}
\affiliation{Max Planck Institute for the Physics of Complex Systems, N\"othnitzer Stra\ss e 38, 01187 Dresden, Germany}
\affiliation{\smash{Max Planck Institute of Molecular Cell Biology and Genetics, Pfotenhauerstra\ss e 108, 01307 Dresden, Germany}}
\affiliation{Center for Systems Biology Dresden, Pfotenhauerstra\ss e 108, 01307 Dresden, Germany}
\author{Raymond E. Goldstein}
\email{r.e.goldstein@damtp.cam.ac.uk}
\affiliation{Department of Applied Mathematics and Theoretical Physics, Centre for Mathematical Sciences, 
\\ University of Cambridge, 
Wilberforce Road, Cambridge CB3 0WA, United Kingdom}
\author{{Diana~Cholakova}}
\email{dc@lcpe.uni-sofia.bg}
\affiliation{Department of Chemical and Pharmaceutical Engineering, Faculty of Chemistry and Pharmacy, \\ University of Sofia, 1 James Bourchier Avenue, 1164 Sofia, Bulgaria}
\author{Nikolai Denkov}
\email{nd@lcpe.uni-sofia.bg}
\affiliation{Department of Chemical and Pharmaceutical Engineering, Faculty of Chemistry and Pharmacy, \\ University of Sofia, 1 James Bourchier Avenue, 1164 Sofia, Bulgaria}
\author{Stoyan K. Smoukov}
\email{s.smoukov@qmul.ac.uk}
\affiliation{School of Engineering and Materials Science, Queen Mary University of London, Mile EndRoad, London E14NS, UK}
\date{\today}%


\maketitle 
Garc\'ia-Aguilar \emph{et al.}~\cite{garcia21} have shown that the deformations of ``shape-shifting droplets'', reported 
in a series of experimental papers spawned by Refs.~\cite{denkov15,guttman16a}, are consistent with an elastic model. Here 
we show that the interplay between surface tension and intrinsic curvature in this model is mathematically
equivalent to a physically very different phase-transition mechanism of the same process described
previously~\cite{haas17,haas19}. Hence, the models cannot distinguish between the two mechanisms, and 
it is not possible to claim that one mechanism underlies the observed phenomena without a 
more detailed comparison of the predictions of both mechanisms with experiments. We suggest 
that the increasing number of seemingly contradictory experimental results 
indicates that the two systems \cite{denkov15,guttman16a} are different. The observed ``shape-shifting'' 
processes are therefore likely to be similar outcomes of two very different physical mechanisms.

Using the notation of Ref.~\cite{garcia21}, we consider a faceted droplet deforming under the interplay of surface tension and bending 
elasticity, with energy
\begin{equation}
E=\iint\!{\left[\gamma_0+2\kappa(H-H_0)^2\right]\,\mathrm{d}A};  \label{eq:EC}  
\end{equation}
we shall justify neglect of the stretching and gravity terms in Eq.~(1) of Ref.~\cite{garcia21} presently. 
If the droplet radius $R$ is much larger than $H_0^{-1}$, then $(H-H_0)^2\approx H_0^2$ everywhere except in a 
neighborhood of characteristic extent $f(\delta)H_0^{-1}$ near the facet edges, in which 
$H\approx H_0$, as in Fig.~1(d) of Ref.~\cite{garcia21}. 
The dimensionless $f(\delta)$ depends on the edge geometry, 
for example through the dihedral angle $\delta$. With these approximations,
\begin{subequations}
\begin{align}
E=\gamma\iint\!{\mathrm{d}A}-2\kappa H_0\int\!{f(\delta)\,\mathrm{d}\ell},     
\end{align}
where $\gamma=\gamma_0+2\kappa H_0^2$, as in Ref.~\cite{garcia21}, and the line integral is along the facet edges. Rescaling lengths with $R$, the scaled energy $\hat{E}=E/\gamma R^2$ is 
\begin{align}
\hat{E}=\iint\!{\mathrm{d}\hat{A}}-\alpha\int{f(\delta)\,\mathrm{d}\hat{\ell}},
\label{2b}
\end{align}
\end{subequations}
with dimensionless tension $\alpha=2\kappa H_0/\gamma R$. In Ref.~\cite{haas19} we obtained the same functional form 
\eqref{2b} for a phase transition model in which deformations are driven by formation of a metastable rotator
phase~\cite{denkov15,haas17,haas19} near the droplet edges. In that case, $\alpha=A\Delta\mu/\gamma R$, 
where $A$ is a characteristic cross-sectional area of rotator phase and $\Delta\mu$ is a difference of chemical potentials. 

To justify neglect of stretching and buoyancy energies, we consider a typical 
droplet radius $R=10\,\mu$m \cite{denkov15}, so 
$R\gg H_0^{-1}\approx 60\,\text{nm}$~\cite{garcia21} and ${r\equiv RH_0\approx 170}$. The relative importance of stretching, 
buoyancy, and intrinsic curvature depends on $r$, the non-dimensional parameters $\Upsilon$, $\Pi$ defined in Ref.~\cite{garcia21}, and the non-dimensional energy differences $\Delta\mathcal{E}_S$, $\Delta\mathcal{E}_G$, 
$\Delta\mathcal{E}_H$ computed from its Table~I. With $\Upsilon\approx 4$, $\Pi\approx 10^{-8}$~\cite{garcia21}, for the icosahedron-platelet transition,
\begin{align}
&\left|\Delta\mathcal{E}_H\right|r\!\approx\! 9400,\;\Upsilon\left|\Delta\mathcal{E}_S\right|r^2\!\approx\! 35,\;\Pi\left|\Delta\mathcal{E}_G\right|r^4\!\approx\!29.
\end{align}
Hence $\left|\Delta\mathcal{E}_H\right|r\gg\Upsilon\left|\Delta\mathcal{E}_S\right|r^2,\Pi\left|\Delta\mathcal{E}_G\right|r^4$; the same separation holds at the sphere-icosahedron transition. Thus, from Eq.~(3) of Ref.~\cite{garcia21}, intrinsic curvature swamps stretching and buoyancy, justifying \eqref{eq:EC}. 

Estimating $\alpha$ reinforces the equivalence: for the elastic mechanism, using Fig.~2(d) of Ref.~\cite{garcia21} to estimate ${\Gamma\approx 0.02}$ at the icosahedron-platelet transition, we find ${\alpha=2(\Gamma r)^{-1}\approx 0.6}$; for the phase transition mechanism, $A\approx 0.3\,\mu$m$^2$ \cite{cholakova19b}, ${\Delta\mu\approx 6\cdot 10^5\,\text{N}/\text{m}^2}$~\cite{haas17}, ${\gamma\approx 5\,\text{mN/m}}$~\cite{denkov16}, so $\alpha\approx 4$. 

The calculations of Garc\'ia-Aguilar \emph{et al.} consider static shapes \cite{garcia21}, and cannot  
show, for example, that an icosahedral droplet would flatten dynamically into a hexagonal platelet rather than 
a different, lower energy shape. Because of the model equivalence, the results of
Ref.~\cite{haas19} showing that an icosahedral droplet can flatten dynamically into a 
hexagonal platelet under the phase-transition mechanism also show it is possible under 
the elastic mechanism of Ref.~\cite{garcia21}.

Experimental studies of ``shape-shifting'' droplets have obtained seemingly contradictory results:  
surface tension measurements~\cite{guttman17,denkov16} differed by orders of magnitude; cryoTEM experiments 
showed monolayers at the 
droplet surface~\cite{guttman19}, while differential scanning calorimetry detected multilayers~\cite{cholakova19b}. However, 
the cationic surfactant $\text{C}_{18}\text{TAB}$ used in Refs.~\cite{guttman16a,guttman17,guttman19} has a relatively high 
surface freezing temperature, while Refs.~\cite{denkov15,haas19,denkov16,cholakova19b} used different  
surfactants covering a 
range of freezing temperatures. These real differences of the 
experimental systems~\cite{denkov16,cholakova19b,cholakova19} and the 
corresponding and mathematically equivalent phase-transition and elastic mechanisms are therefore physically different 
realizations of a more general ``shape-shifting'' mechanism based on the interplay of positive surface tension and 
negative edge tension in 
faceted droplets.

\begin{acknowledgments}
We thank L. Giomi for discussions. This work was supported in part by the Max Planck Society (P.A.H.) and Fellowships EP/M017982/1 (R.E.G.) and EP/R028915/1 (S.K.S.) from the Engineering and Physical Sciences Research Council.
\end{acknowledgments}

\bibliography{comment}
\end{document}